\documentclass[prl,twocolumn,showpacs,superscriptaddress,floatfix, nofootinbib]{revtex4-2}
\usepackage[caption=false]{subfig}
\usepackage{amsmath,graphicx}
\usepackage{amsmath,amssymb,mathrsfs,esint}
\usepackage{multirow}
\usepackage{algorithm}
\usepackage[noend]{algpseudocode}
\usepackage{comment}
\usepackage{tabularx}
\usepackage{bm}
\usepackage[normalem]{ulem}
\usepackage[bookmarks=true,
   colorlinks=true,
   linkcolor=blue,
   urlcolor=blue,
   citecolor=blue,
   bookmarks=true,
   hyperindex=true
]{hyperref}
\usepackage{physics}
\usepackage{xifthen}
\usepackage{tikz, adjustbox}
\usetikzlibrary{quantikz2}

\newcommand{\mixing}[1][]{\ifthenelse{\isempty{#1}}{\vb*{\beta}}{\beta_{#1}}}
\newcommand{\problem}[1][]{\ifthenelse{\isempty{#1}}{\vb*{\gamma}}{\gamma_{#1}}}

\begin{document}

\title{Experimental factoring integers using fixed-point-QAOA \\ with a trapped-ion quantum processor}

\author{Ilia V. Zalivako}
\affiliation{P.N. Lebedev Physical Institute of the Russian Academy of Sciences, Moscow 119991, Russia}
\affiliation{Russian Quantum Center, Skolkovo, Moscow 121205, Russia}

\author{Andrey Yu. Chernyavskiy}
\affiliation{Russian Quantum Center, Skolkovo, Moscow 121205, Russia}

\author{Anastasiia S. Nikolaeva}
\affiliation{P.N. Lebedev Physical Institute of the Russian Academy of Sciences, Moscow 119991, Russia}
\affiliation{Russian Quantum Center, Skolkovo, Moscow 121205, Russia}
\affiliation{National University of Science and Technology ``MISIS”,  Moscow 119049, Russia}

\author{Alexander S. Borisenko}
\affiliation{P.N. Lebedev Physical Institute of the Russian Academy of Sciences, Moscow 119991, Russia}
\affiliation{Russian Quantum Center, Skolkovo, Moscow 121205, Russia}

\author{Nikita V. Semenin}
\affiliation{P.N. Lebedev Physical Institute of the Russian Academy of Sciences, Moscow 119991, Russia}
\affiliation{Russian Quantum Center, Skolkovo, Moscow 121205, Russia}

\author{Kristina P. Galstyan}
\affiliation{P.N. Lebedev Physical Institute of the Russian Academy of Sciences, Moscow 119991, Russia}
\affiliation{Russian Quantum Center, Skolkovo, Moscow 121205, Russia}

\author{Andrey E. Korolkov}
\affiliation{P.N. Lebedev Physical Institute of the Russian Academy of Sciences, Moscow 119991, Russia}
\affiliation{Russian Quantum Center, Skolkovo, Moscow 121205, Russia}

\author{Sergey V. Grebnev}
\affiliation{Russian Quantum Center, Skolkovo, Moscow 121205, Russia}

\author{Evgeniy O. Kiktenko}
\affiliation{Russian Quantum Center, Skolkovo, Moscow 121205, Russia}
\affiliation{National University of Science and Technology ``MISIS”,  Moscow 119049, Russia}

\author{Ksenia Yu. Khabarova}
\affiliation{P.N. Lebedev Physical Institute of the Russian Academy of Sciences, Moscow 119991, Russia}
\affiliation{Russian Quantum Center, Skolkovo, Moscow 121205, Russia}

\author{Aleksey K. Fedorov}
\affiliation{P.N. Lebedev Physical Institute of the Russian Academy of Sciences, Moscow 119991, Russia}
\affiliation{Russian Quantum Center, Skolkovo, Moscow 121205, Russia}
\affiliation{National University of Science and Technology ``MISIS”,  Moscow 119049, Russia}

\author{Ilya A. Semerikov}
\affiliation{P.N. Lebedev Physical Institute of the Russian Academy of Sciences, Moscow 119991, Russia}
\affiliation{Russian Quantum Center, Skolkovo, Moscow 121205, Russia}

\author{Nikolay N. Kolachevsky}
\affiliation{P.N. Lebedev Physical Institute of the Russian Academy of Sciences, Moscow 119991, Russia}
\affiliation{Russian Quantum Center, Skolkovo, Moscow 121205, Russia}

\begin{abstract}
Factoring integers is considered as a computationally-hard problem for classical methods, whereas there exists polynomial-time Shor's quantum algorithm for solving this task. 
However, requirements for running the Shor's algorithm for realistic tasks, which are beyond the capabilities of existing and upcoming generations of quantum computing devices,  
motivates to search for alternative approaches.
In this work, we experimentally demonstrate  factoring of the integer with a trapped ion quantum processor using the Schnorr approach and a modified version of quantum approximate optimization algorithm (QAOA).
The key difference of our approach in comparison with the recently proposed QAOA-based factoring method is the use of the fixed-point feature, which relies on the use of universal parameters.
We present experimental results on factoring $1591=37\times43$ using 6 qubits as well as simulation results for $74425657=9521\times7817$ with 10 qubits and $35183361263263=4194191\times8388593$ with 15 qubits. 
Alongside, we present all the necessary details for reproducing our results and analysis of the performance of the factoring method, the scalability of this approach both in classical and quantum domain still requires further studies. 
\end{abstract}

\maketitle

{\it Introduction}.
Shor's algorithm~\cite{Shor1994,Shor1999} for factoring integers has become one of the examples of a practically relevant problem, which is hard for classical computer yet amenable for quantum processors. 
The implication of the integer factorization problem to the widely adopted cryptographic schemes, such as the RSA cryptosystem~\cite{RSA1978}, 
is a clear motivation for studying its practical complexity within both the classical and quantum approaches~\cite{Bernstein2017,Fedorov2021-2}. 
Proof-of-concept experimental factoring of 15, 21, and 35 have been demonstrated on superconducting~\cite{Martinis2012-2}, trapped ion~\cite{Blatt2016}, and photonic~\cite{Pan2007,White2007,OBrien2012} quantum computers. 
However, the implementation of Shor's algorithm for breaking of actually employed cryptosystems requires resources, which seem to be far beyond the capabilities of existing and upcoming generations of quantum computing devices.
For example, in order to factor a 2048-bit RSA integers (that is, an integer $N=pq$ where $p,q$ are distinct primes) one needs 8 hours using 20 million noisy qubits~\cite{Gidney2021}). 
Various approaches to implement factoring with fewer resources~\cite{Coppersmith2002,Bocharov2017,Regev2024} or even with existing noisy intermediate-scale quantum (NISQ) devices~\cite{Aspuru-Guzik2019,Peng2019,Wang2020} are under development.
Recent proposal~\cite{Yan2022} claims a possibility of solving the factorization problem with sublinear quantum resources.
This approach is conceptually similar to the idea of variational quantum factoring~\cite{Aspuru-Guzik2019}, where quantum approximate optimization algorithm (QAOA)~\cite{Farhi2014,Farhi2019} is used.
In contrast to the original Shor's algorithm, QAOA can be efficiently run on NISQ devices~\cite{Monroe2020,Babbush2021,Bharti2021,Babbush2021-4}.
Such variational approach has been used before in experiments~\cite{Cao2021} on factoring 1099551473989, 3127, and 6557 with 3, 4, and 5 qubits, correspondingly; however, such an approach requires further analysis of scalability.  
In Ref.~\cite{Yan2022} a theoretical path towards factoring RSA 2048-bit key with 372 physical qubits only has been declared.
The proposed method uses several steps of the lattice reduction-based Schnorr's factorization technique~\cite{Schnorr2021} where QAOA~\cite{Farhi2014,Farhi2019} is employed to reduce a number of iterations required to factorize the number. 
However, as it has been shown~\cite{Fedorov2023,Khattar2023} such an approach encounters a number of pitfalls coming both from classical and quantum domains. 

In this work, we demonstrate that certain obstacles of the QAOA-based factoring can be overcame by switching to an original fixed-point version of QAOA~\cite{chernyavskiy2023fixed}.
While it became a routine to run QAOA with classical optimization of expectation values with respect to the parameters, such an approach suffers from the problem of global optimization and statistical fluctuations. To our knowledge, the alternative idea to exploit so-called universal angles (parameters) in QAOA has been presented for the first time in Ref.~\cite{brandao2018fixed}. 
We follow the latter approach so that in our fixed-point version of QAOA~\cite{chernyavskiy2023fixed} we use fixed optimal parameters from the corresponding training set of tasks, normalize it (i.e., Hamiltonians), 
and then search for angles providing the maximum minimal increase in the probability of a correct answer, whereas the Max-Min problem is solved via evolution optimization.
This allows us to solve reliably the closest vector problem, which lies in the basis of the Schnorr's algorithm, with the use of the quantum device. 
Within this approach we demonstrate experimental factoring of the number $1591=37\times43$ using 6 qubits with a trapped ion quantum processor. 
We also present simulation results for $74425657=9521\times7817$ and $35183361263263=4194191\times8388593$ with 10 and 15 qubits, correspondingly.
Although we expect that one of the difficulties in the realization of the QAOA-based factoring is resolved, still this approach requires further scalability studies.

{\it Fixed-point QAOA-based factoring}.
The crucial component of the Schnorr's factoring algorithm is the search for smooth relation pairs of integers, so-called sr-pairs.
As soon as we have a sufficient number of such pairs, which is larger than the size of the factoring base (hyperparameter of the algorithm), we can form a system of linear equations that appears to be degenerate and always has a solution. This solution, by a classical Fermat's method (see, for example, Ref.~\cite{Fermat}), provides a factorization with a high probability.
The problem of sr-pairs search can be reduced to the closest vector problem (CVP) on a lattice. 
The closer the found solution to the desired vector, the greater the chance of obtaining a smooth relationship. 
Schnorr's method relies on solving the CVP with the classic approximate LLL-reduction (Lenstra–Lenstra–Lovász) algorithm~\cite{lenstra1982factoring}. As this algorithm gives only an approximate real-valued solution, in the original paper by Schnorr it was rounded to the closest integer value at the last step. 
The idea behind the recent proposal~\cite{Yan2022} is to choose the rounding side for each variable to find the closest integer-valued solution, which in turn reduces to a quadratic unconstrained binary optimization (QUBO) problem.
Such class of problems is amenable to solving with QAOA.

QAOA is based on the trotterization of the adiabatic evolution of the following form:
\begin{equation}\label{eq:qaoa_circ}
    \begin{aligned}
        &\ket{\mixing,\problem}=
    U(\mixing[p], \problem[p])\ldots
    U(\mixing[1], \problem[1])
    \ket{+}^{\otimes n}, \\
        &U(\mixing[j], \problem[j]) = e^{-i \mixing[j] H_M}e^{-i \problem[j] H_P}
    \end{aligned}
\end{equation}
where $\mixing=\{\mixing[j]\}$ and $\problem=\{\problem[j]\}$ are circuit parameters (angles), hyperparameter $p$ is the number of layers, $\ket{+}$ is the $+1$ eigenstate of $\sigma_x$ Pauli matrix, $H_M=\sum_{k}{\sigma_x^{(k)}}$ is the mixing Hamiltonian (here $\sigma_x^{(k)}$ is $\sigma_x$ acting on $k$th qubit), and $H_P$ is the problem Hamiltonian, which in most cases directly encodes the Ising form of a QUBO problem to be solved.

\newcommand{\zang}{\theta}
\newcommand{\xang}{2\beta}






\begin{figure}
	\centering
	\begin{adjustbox}{width=0.99\linewidth}\Large
		\begin{quantikz}
			\lstick{$\ket{0}_1$}&\gate{R_y(\frac{\pi}{2})}&\ctrl{1}&\ctrl{2}&\ctrl{3}&\ctrl{4}&\ctrl{5}&&&&&&&&&&&\gate{R_z(\zang_1)}&\gate{R_x(\xang)}&\meter{}\\
			\lstick{$\ket{0}_2$}&\gate{R_y(\frac{\pi}{2})}&\ctrl{0}&&&&&\ctrl{1}&\ctrl{2}&\ctrl{3}&\ctrl{4}&&&&&&&\gate{R_z(\zang_2)}&\gate{R_x(\xang)}&\meter{}\\
			\lstick{$\ket{0}_3$}&\gate{R_y(\frac{\pi}{2})}&&\ctrl{0}&&&&\ctrl{0}&&&&\ctrl{1}&\ctrl{2}&\ctrl{3}&&&&\gate{R_z(\zang_3)}&\gate{R_x(\xang)}&\meter{}\\
			\lstick{$\ket{0}_4$}&\gate{R_y(\frac{\pi}{2})}&&&\ctrl{0}&&&&\ctrl{0}&&&\ctrl{0}&&&\ctrl{1}&\ctrl{2}&&\gate{R_z(\zang_4)}&\gate{R_x(\xang)}&\meter{}\\
			\lstick{$\ket{0}_5$}&\gate{R_y(\frac{\pi}{2})}&&&&\ctrl{0}&&&&\ctrl{0}&&&\ctrl{0}&&\ctrl{0}&&\ctrl{1}&\gate{R_z(\zang_5)}&\gate{R_x(\xang)}&\meter{}\\
			\lstick{$\ket{0}_6$}&\gate{R_y(\frac{\pi}{2})}&&&&&\ctrl{0}&&&&\ctrl{0}&&&\ctrl{0}&&\ctrl{0}&\ctrl{0}&\gate{R_z(\zang_6)}&\gate{R_x(\xang)}&\meter{}\\
		\end{quantikz}
	\end{adjustbox}
	\caption{Architecture of the executed quantum circuits in fixed-point-QAOA algorithm. Each pair of connected black circles corresponds to $ZZ(\chi_{ij})$ gate acting on $i$-th and $j$-th qubits, where for each involved qubit pair $\chi_{ij}$ is unique. Angles $\zang_i$ in $R_z(\zang_i)$ gates are also different for each $i$-th qubit in each circuit. $\beta$ in $R_x(\xang)$ is equal to 2.64.
	For each of 9 executed circuits parameters of these gates are given in table \ref{tab:angles} of Supplementary Materials.
	}
	\label{fig:6qb-circ}
\end{figure}
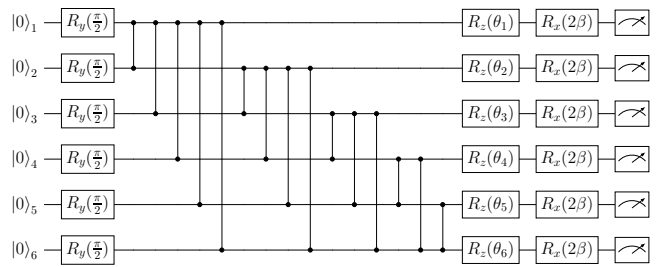
The most common approach to QAOA is to classically optimize the expectation value $E(\mixing,\problem) = \expval{H_P}{\mixing,\problem}$ being estimated by the set of measurements (shots) on a quantum processor (see, e.g.,  Refs.~\cite{guerreschi2019qaoa, zhou2020quantum, fernandez2022study}). In contrast, in the seminal QAOA paper~\cite{Farhi2014} relies on searching optimal angles utilizing the efficient exact classical calculation of $E$ (which was presented for Max-Cut problems on 3-regular graphs~\cite{Farhi2014}) followed by sampling on a quantum processor. We have used an alternative approach based on the empirical hypothesis of close optimal angles for different instances of the same problem type~\cite{brandao2018fixed, galda2021transferability, wurtz2021fixed}.

To find fixed QAOA parameters, we use the training set consisting of 100 QUBO subproblems arised during factoring $N=48567227$ on $n=10$ qubits. As optimal QAOA problem angles $\problem$ scale together with QUBO coefficients, we normalize every QUBO coefficient matrix by its maximal value~\cite{chernyavskiy2023fixed}. The ratio $P_q/P_c$ of the probability $P_q$ to measure the optimal (minimal) answer to its classical random sampling counterpart $P_c$ was used as an optimization metric, and its minimum over the training set was maximized using random mutations optimization algorithm~\cite{chernyavskiy2013calculation, bantysh2020quantum}. To minimize the quantum circuit depth, we use just a single layer of QAOA ($p=1$), which significantly increase robustness of the quantum part of the algorithm.
The quantum circuit for a single layer of QAOA used in the algorithm has the form presented in Fig.~\ref{fig:6qb-circ}.
The resulting single-layer QAOA parameters used in the factorization are $\gamma_1:=\gamma_*=2.64$ and $\beta_1:=\beta_*=0.33.$
The fixed-parameters approach allows avoiding the classical-quantum hybrid optimization procedure and fits well with the demands of Schnorr's method: one does not need to obtain the exact or suboptimal solution of CVP, but sample solutions close to the target vector to increase the probability of forming a set of sr-pairs. 

In the classical part of the algorithm, we directly follow Refs.~\cite{Yan2022} and~\cite{Fedorov2023}. We use the main factor base of the size $B_1=6$ (which is equal to the number of qubits), the relaxed factor base size for sr-pairs verification is $B_2=11,$ the rounding parameter of lattice/target formation procedure is $c=1.5$ and the parameter of LLL-reduction is $\delta=0.75$. For each lattice (which is formed by a random permutation of the diagonal), we conduct $5$ measurements (shots) of each circuit. Due to a strongly stochastic nature of the algorithm the required number of circuits varies. The details of a single run of the factorization algorithm including the exact form of the circuit and corresponding parameters are provided in the Supplemental Material.

\begin{figure*}
     \subfloat[$N=43\times 37,\;n=6, B2=11,\;c=1.5,\;N_{sh}=5$]{\includegraphics[width=0.33\textwidth]{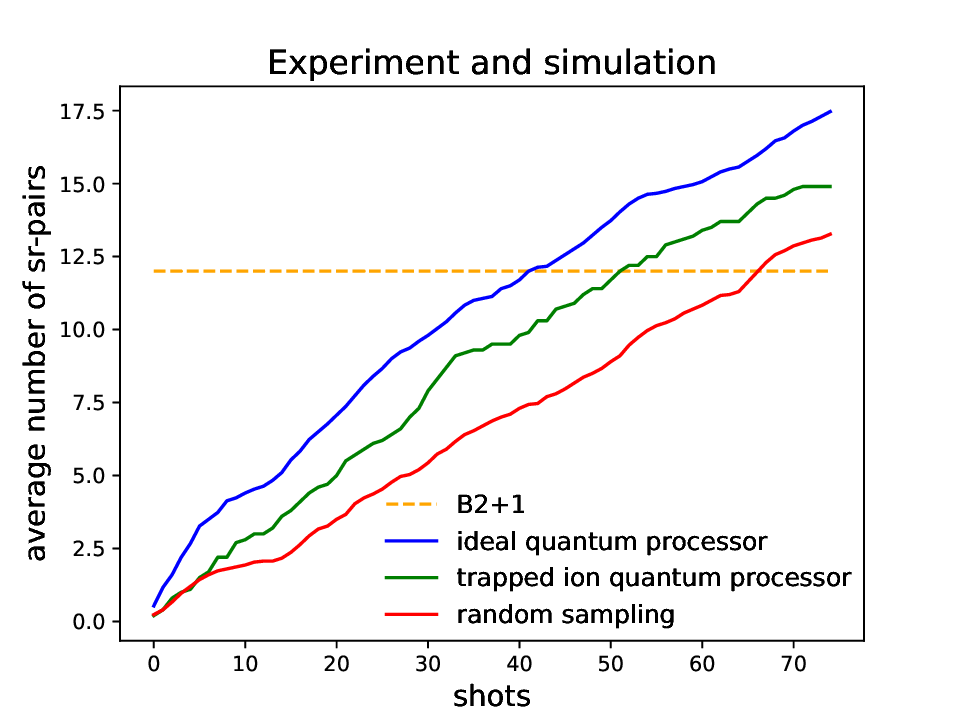}    
    \label{fig:sr-pairs-a}}
     \subfloat[$N=9521\times 7817,\;n=10,B2=50,\;c=4,\;N_{sh}=20$]{\includegraphics[width=0.33\textwidth]{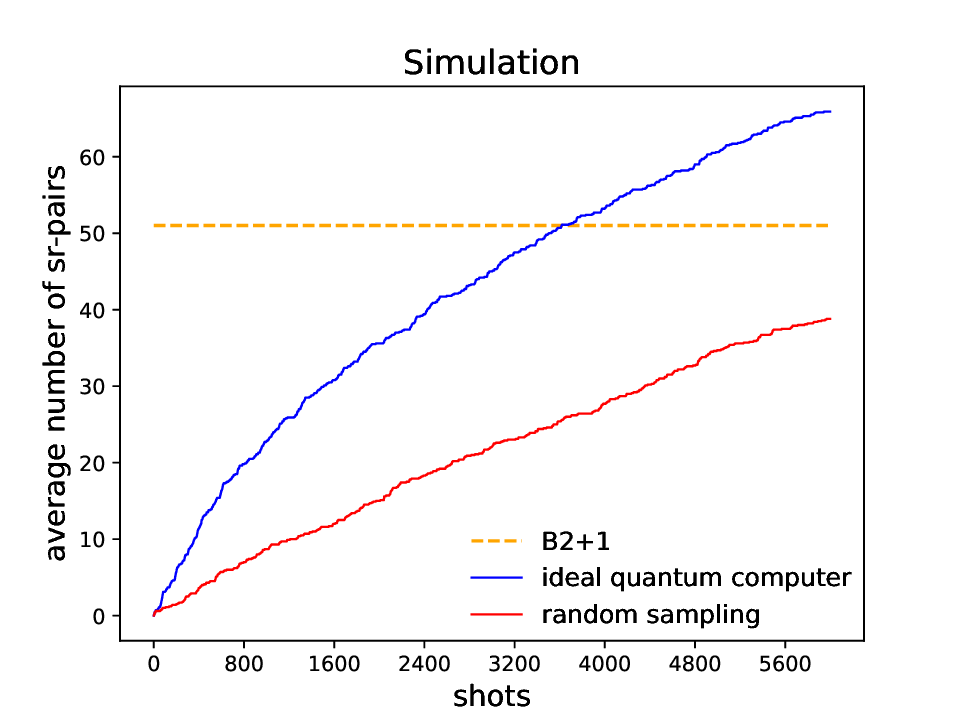}}
    \subfloat[$N=4194191\times 8388593,\;n=15,B2=450,\;c=4,\;N_{sh}=1000$]{\includegraphics[width=0.33\textwidth]{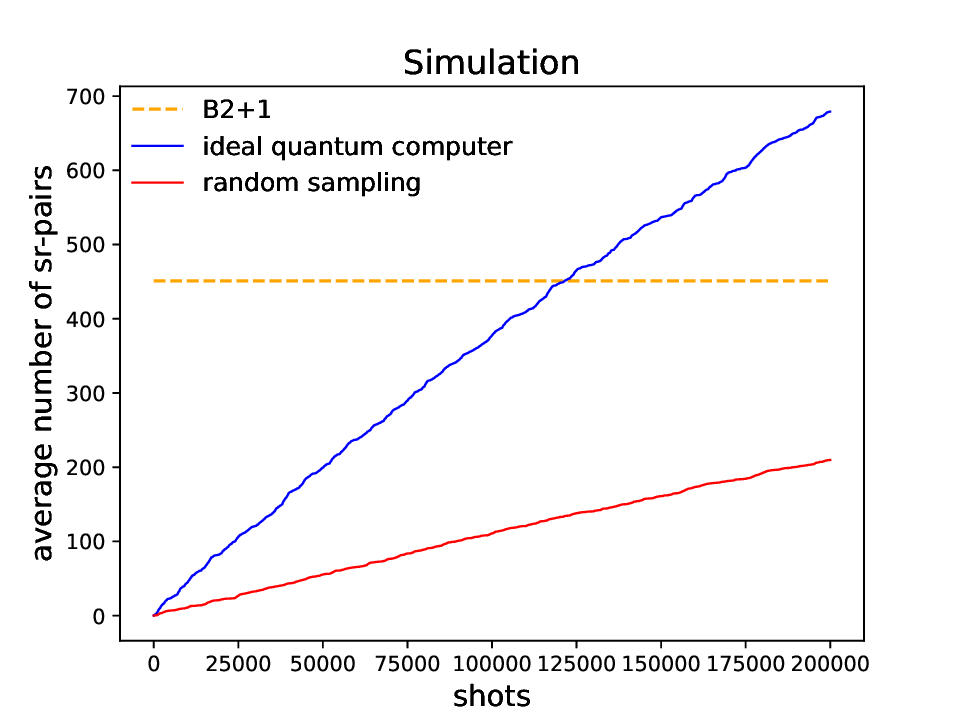}}
    \centering

    \caption{
    A comparison of sr-pairs collection rates between cases where QUBO-subproblems samples are generated with a random sampling (red lines), a noiseless quantum emulator (blue) and a real trapped-ion quantum processor (green) for different number of qubits. The left sub-figure shows both experimental (averaged over 10 runs) and simulation data (averaged over 30 runs), while other figures contain only simulation results (averaged over 10 trajectories). Here $N$ stands for the factorized number, $n$ is for the number of qubits, and $N_{sh}$ is for a number of shots per circuit. The dashed horizontal line shows a $B2+1$ sr-pairs threshold which guarantees the factorization.
    }
    \label{fig:sr-pairs}
\end{figure*}

{\it Experimental setup}.
Experimental demonstration of the algorithm was performed with a quantum processor based on a chain of ten ultracold $^{171}$Yb$^{+}$ ions in a linear Paul trap. Details of the setup can be found in Refs.~\cite{kazmina2023demonstration, zalivako2024}. Qubits are encoded in states $|0\rangle = \,^2S_{1/2}(F=0, m_F=0)$ and $|1\rangle = \,^2D_{3/2}(F=2, m_F=0)$, coupled by an optical E2 transition at wavelength $\lambda = 435.5$~nm. 
While the setup supports usage of all five Zeeman sublevels of the upper state for the information encoding (i.e. we have the qudit processor~\cite{zalivako2024}); in this work we have used the processor in the qubit regime.

Before each experimental shot ions are Doppler cooled to the temperatures of approximately 1.5 $mK$, which is followed by the sideband-cooling of all radial motional modes close to the ground state and initialization to the $|0\rangle$ state by optical pumping~\cite{zalivako2024}. On the next stage the target native gates sequence is being implemented. In our system single-qubit native gates are $R_{\phi}(\theta) = \exp(-i\sigma_{\phi}\theta/2)$ and $R_z(\theta)=\exp\left(i\theta|1\rangle\langle1|\right)$, where $\sigma_\phi = \cos\phi \sigma_x + \sin\phi \sigma_y$, and $\phi, \theta$ --- arbitrary angles. The first operation is performed by applying a laser pulse, resonant to the $|0\rangle\to|1\rangle$ transition. In this case $\phi$ is determined by the relative phase of the laser field and the qubit, while $\theta$ is determined by the pulse duration. The $R_{z}(\theta)$ is a virtual gate~\cite{McKay2017} and is performed by shifting phases of all successive laser pulses applied to this ion. A native two-qubit operation for this system is a M\o{}lmer-S\o{}rensen gate ~\cite{Blatt2003-2,Molmer-Sorensen1999,Molmer-Sorensen1999-2,Molmer-Sorensen2000} $R_{xx}(2\chi) \equiv XX(\chi)=\exp\left(-i\chi\sigma_x\otimes\sigma_x\right)$. This gate is implemented by illuminating a target pair of ions with a bichromatic laser fields, coupling their electronic states with a collective motional degrees of freedom (in our case we use radial motional modes). These common motional modes serve as mediator, coupling both qubits. The laser fields are amplitude-modulated to decouple all electronic degrees of freedom from motional ones at the end of the gate and reduce sensitivity of the operation to the experimental parameters~\cite{choi2014optimal}. The processor supports $XX(\chi)$ gates with arbitrary $\chi$ and all-to-all connectivity. We also include $R_{zz}(2\chi) \equiv ZZ(\chi)=\exp\left(-i\chi\sigma_z\otimes\sigma_z\right)$ gate in the list of supported operations, which is automatically hardware-efficiently transpiled as $ZZ(\chi) = (R_{y}(\pi/2)\otimes R_{y}(\pi/2)) XX(\chi) (R_{y}(-\pi/2)\otimes R_{y}(-\pi/2))$ in the processor. At the end of each experimental shot the quantum register readout is performed using electron-shelving technique on the $\ket{^{2}S_{1/2}} \to \ket{^{2}P_{1/2}}$ transition at 369~nm~\cite{semenin2021optimization, zalivako2024}. Ions fluorescence in this process is collected with a high numeric aperture lens and is sent via an array of multimode fibers to the multichannel photomultiplier tube.

Fidelities of the single-qubit and two-qubit operations are 99.95\% and 95\%, which are measured using randomized benchmarking~\cite{magesan2012characterizing}, and parity oscillations observation~\cite{Blatt2008}, correspondingly. The qubits coherence time $T^*_2 = 30$~ms was extracted from decay of Ramsey fringes contrast with increasing delay between $\pi/2$ pulses.
To reduce cross-talk during single-qubit operations all $R_{\phi}(\theta)$ gates in the circuits are substituted with their composite analogues using SK1 scheme~\cite{brown2004arbitrarily}. Particularly, two $2\pi$ rotations around specific axes are added after each single-qubit gate, which are known to suppress both cross-talks and rotation angle fluctuations. 

{\it Experimental results}.
In the experiment, we use the Schnorr's approach assisted with the fixed-angles QAOA to factorize number $1591=37\times43$ using 6 qubits. 

In a single sample run of the experiment (for details, see Supplemental Material), $B_2+1$ sr-pairs required to deterministically factorize the number were found in $43$ steps (shots) using $9$ different quantum circuits (each circuit repeated $5$ times followed by the next circuit). However, in this particular sample run the first $39$ shots appeared to be already sufficient to factorize the number. 
We have compared the average speed of collecting unique sr-pairs in three cases: (i) random sampling; (ii) experimentally obtained samples; (iii) samples obtained with noiseless emulator (see Fig.~\ref{fig:sr-pairs-a}). The figure demonstrates the advantage of the quantum processor sampling results over the random sampling. However, the presence of the noise in the system decreases the efficiency of the method in comparison with a noiseless emulator. To illustrate the level of the noise in the quantum processor we also measured the output states probability distributions for several used circuits with better averaging and compared it with results expected in the absence of errors (for details, see Supplemental Material).

In this experiment we chose to use 6 qubits as a trade-off between the problem size and quantum circuits fidelity. Numeric simulations show, that the expected advantage over random sampling in QUBO-subproblems increases with the growth of qubits number and magnitude of a number to factorize (e.g. see Fig. \ref{fig:sr-pairs}). At the same time as the number of two-qubit operations in each circuit is equal to $n(n-1)/2$, where $n$ is a number of qubits, the quantum sampling fidelity decreases with larger $n$. In our experiments $n=6$ was the smallest number of qubits, where the advantage over random sampling was observed experimentally despite the better sampling fidelity at $n < 6$.

A number of shots per circuit was chosen using numerical simulations. It was set to be sufficient to find enough sr-pairs, keeping the total number of shots minimal.

{\it Scalability analysis}.
The initial complexity estimates presented in Schnorr's work did not lead to practical results for factoring large numbers, however, the effectiveness of the method has still neither been proven nor strictly disproved. 
Based on Refs.~\cite{Khattar2023, aboumrad2023quantum, luan2023lattice,Fedorov2023} and own numerical experiments, the following difficulty can be noted: the probability that estimates obtaining an sr-pair by suboptimal solutions of CVP problem (obtained by classical or quantum methods) does not directly lead to the probability of obtaining a set of \textit{unique} sr-pairs needed for the factorization. 
Such analysis is also complicated by a large set of hyperparameters. 
Thus, the presented approach need further research on factorization speed and hyperparameters influence.  
At least it is important to compare the approach with classical methods other than uniform random sampling, including quantum-inspired techniques (e.g. \cite{tesoro2024}).

{\it Conclusion and outlook}.
We have considered Schnorr factoring scheme, where following the idea from Ref.~\cite{Yan2022} we adopt QAOA method at the last step of Babai's algorithm. 
However, for the first time we used a fixed-point feature of QAOA \cite{chernyavskiy2023fixed, brandao2018fixed} for the factoring problem and were able to factor a specific integer.  
To the best of our knowledge, it is the first successful experimental factoring of a particular integer with fixed-point QAOA-assisted Schnorr approach, whereas previously it was only experimentally presented how to obtain some sr-pairs for this task using quantum computers.

To confirm both the overall scheme and the fixed-point approach we experimentally factor $1591=37\times43$ using 6 qubits of the 10-qubit trapped-ion processor.
We have also presented simulation results for $74425657=9521\times7817$ and $35183361263263=4194191\times8388593$ with 10 and 15 qubits, correspondingly.

For further research we leave the questions of the algorithm's efficiency and thorough comparison with classical methods, as well as a more detailed investigation of the quantum processor noise influence.

{\it Note added}. After completion of this work, we became aware of Ref. \cite{priestley2025}, which also suggests using fixed-point QAOA in the same context. Authors have presented an alternative approach of fixed angles search and scaling, and conducted a thorough numerical analysis of QAOA-augmented refinement of CVP problem. In contrast, in our work we consider the complete factorization algorithm and its experimental trapped-ion implementation.

{\it Acknowledgements}.
A.S.N., E.O.K. and A.K.F. acknowledge support from the Priority 2030 program at the NIST ``MISIS'' under the project K1-2022-027.
The experimental part of this work was supported by the Russian Roadmap on Quantum Computing (Contract No. 868-1.3-15/15-2021, October 5, 2021).

\bibliography{bibliography.bib}


\begin{widetext}
\clearpage
~

\section*{Supplemental Material}
In this supplemental section we provide the details of a single run of the factoring algorithm. Let's consider the first random permutation $(1, 3, 2, 5, 6, 4)$ used in the algorithm. 
The corresponding CVP is defined by the lattice
$$L=\begin{pmatrix}
  1  &  0  &  0  &  0  &  0  &  0 \\
  0  &  2  &  0  &  0  &  0  &  0 \\
  0  &  0  &  1  &  0  &  0  &  0 \\
  0  &  0  &  0  &  3  &  0  &  0 \\
  0  &  0  &  0  &  0  &  3  &  0 \\
  0  &  0  &  0  &  0  &  0  &  2 \\
  22  &  35  &  51  &  62  &  76  &  81 
\end{pmatrix}$$
and the target vector
$$t=\begin{pmatrix}
  0  &  0  &  0  &  0  &  0  &  0  &  233 
\end{pmatrix}.$$
The approximate solution given by the Babai's algorithm based on LLL-reduction is
$$\begin{pmatrix}
  19  & -23  & -41  & -32  &  32  &  0 
\end{pmatrix},$$
all elements were rounded \emph{up} to the nearest integer.
The corresponding normalized (to the maximal value) matrix of QUBO coefficients (rounded to $10^{-3}$) is
$$Q=\begin{pmatrix}
 -0.929 & -0.286 &  0.143 &  0.071 &  0.143 &  0.286\\
  0.000 &  1.000 & -0.286 &  0.143 & -0.286 & -0.571\\
  0.000 &  0.000 & -1.643 & -0.286 &  0.643 &  0.071\\
  0.000 &  0.000 &  0.000 & -0.143 &  0.000 & -0.429\\
  0.000 &  0.000 &  0.000 &  0.000 & -2.571 &  0.643\\
  0.000 &  0.000 &  0.000 &  0.000 &  0.000 & -1.429
\end{pmatrix}.$$

\section{Fixed-point-QAOA circuits}

To factor $1591=37\times43$ with fixed-point QAOA we implemented 9 quantum 6-qubit quantum circuits on a trapped-ion processor. Due to the fixed-point feature there is no need in  classical-quantum hybrid optimization, therefore all necessary parameters for circuit construction can be obtained before execution on a quantum hardware. 
Exact architecture of executed quantum circuits with native for the processor single-qubit and two-qubit gates is presented in Fig. \ref{fig:6qb-circ}.
Parameters of the circuits, which correspond to angles in the gates $R_z(\theta_i)$ and $ZZ(\chi_{ij})$, are given in Tab. \ref{tab:angles}.
When $\chi_{ij}$ is equal to zero, $ZZ(\chi_{ij})$ is not implemented. 
We note that to get sufficient statistics it was enough to perform 5 shots for each circuit.
In total, 45 experimental shots were executed on a trapped-ion processor.
To collect 12 sr-pairs 43 shots were enough. 

\section{Experimental QUBO sampling accuracy}
In this section we present comparison between experimentally obtained output states probabilities for circuits 1 and 6 from the Table~\ref{tab:angles} and ones calculated on a noiseless emulator (Fig.~\ref{fig:histograms}).

\begin{figure}
    \centering
     \includegraphics[width=0.9\linewidth]{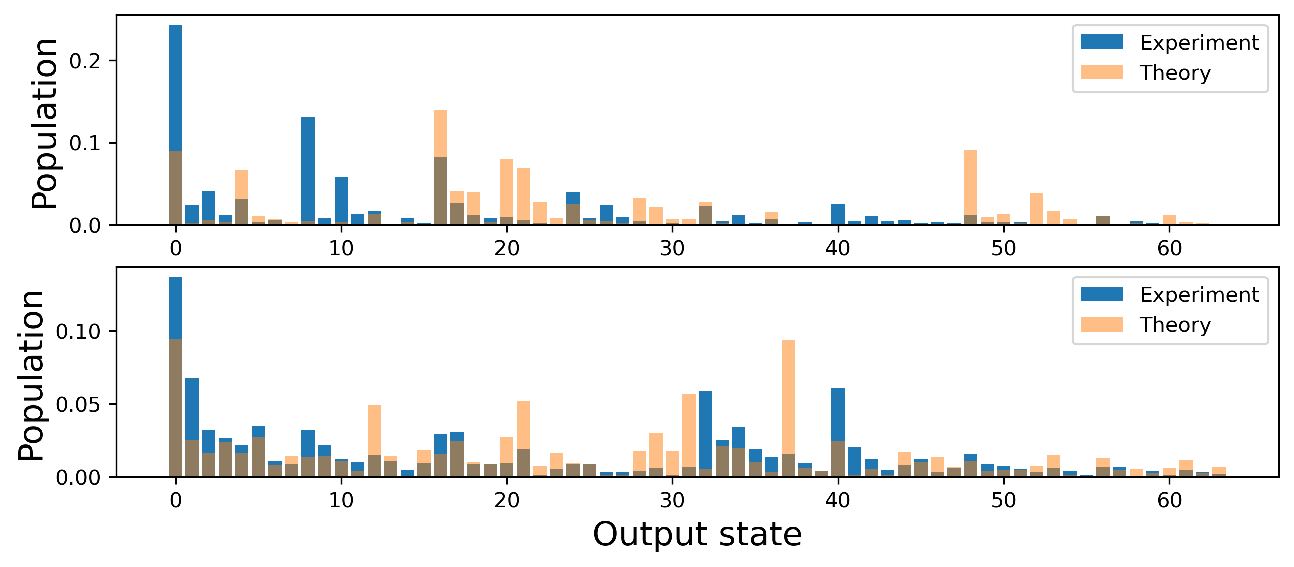}
       
    \caption{Output states probabilities for circuits 1 and 6 from the Table~\ref{tab:angles} sampled by the quantum processor and the noiseless emulator. Output states are numbered as a decimal representation of the output bitstrings. The first qubit corresponds to the high-order digit in the bistrings. Each histogram is an average of 2000 shots.}
    \label{fig:histograms}
\end{figure}

In the Fig.~\ref{fig:histograms5} we also show an analogous output probability distributions for the circuits where we use only 5 qubits to factorize number 437. It can be seen, that the sampling fidelity is generally higher than for a 6 qubit case due to smaller circuit depth. However, for such a small problem size no quantum advantage over random sampling was observed.

\begin{figure}
    \centering
     \includegraphics[width=0.9\linewidth]{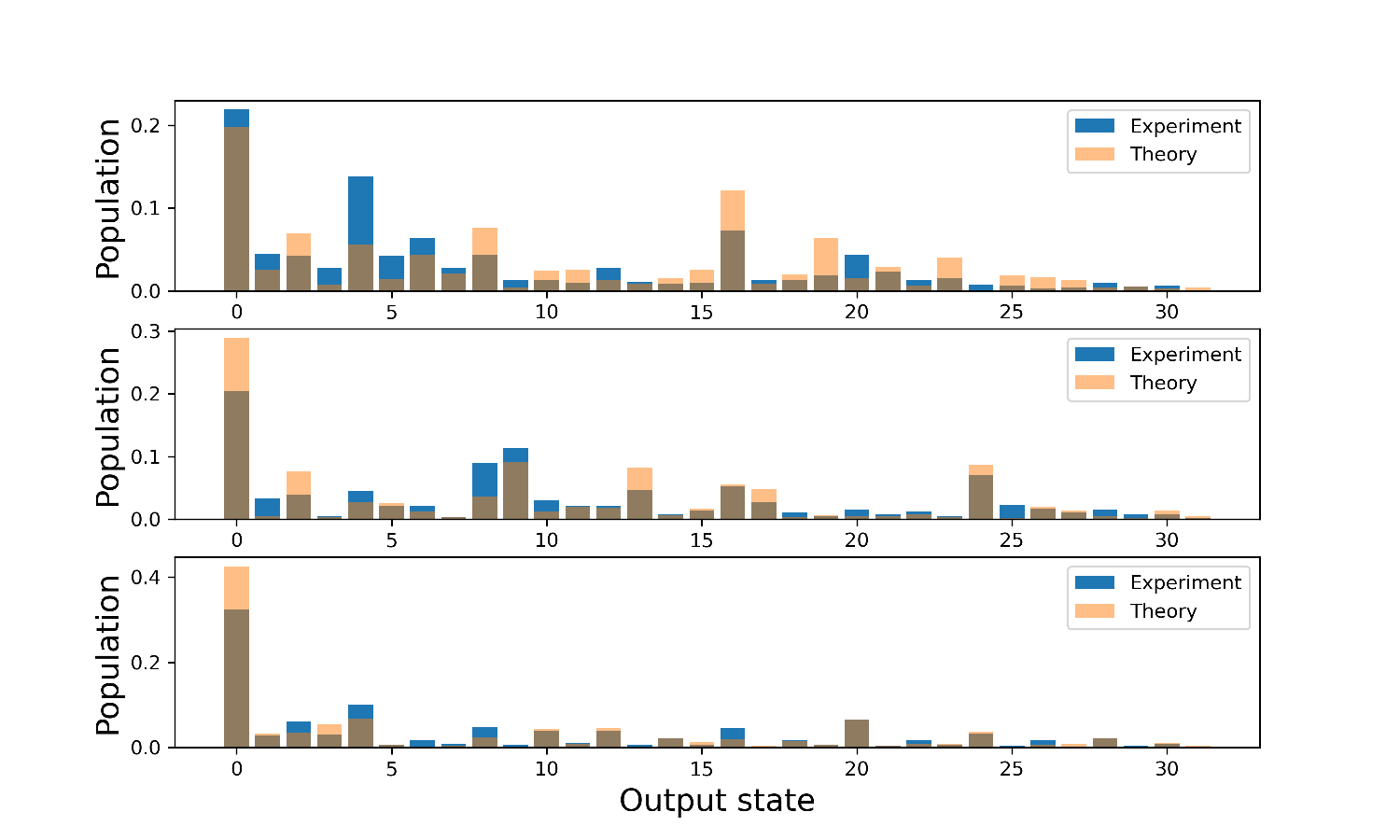}
       
    \caption{Output states probabilities sampled by the quantum processor and the noiseless emulator for a set of circuits used to factorize number 437 using 5 qubits. Output states are numbered as a decimal representation of the output bitstrings. The first qubit corresponds to the high-order digit in the bistrings. Each histogram is an average of 2000 shots.}
    \label{fig:histograms5}
\end{figure}

\begin{table*}[t]
\begin{center}
\begin{tabular}{|c|c|c|c|c|c|c|}
\hline
Step & Permutation & Circuit & Measurement Result & sr-pair & \#sr-pairs & Factoring\\
\hline
1 & (1, 3, 2, 5, 6, 4) & 1 & 010001 &  & 0 & \\
2 & (1, 3, 2, 5, 6, 4) & 1 & 101000 &  & 0 & \\
3 & (1, 3, 2, 5, 6, 4) & 1 & 000100 &  & 0 & \\
4 & (1, 3, 2, 5, 6, 4) & 1 & 001010 &  & 0 & \\
5 & (1, 3, 2, 5, 6, 4) & 1 & 000001 &  & 0 & \\
6 & (4, 1, 3, 6, 5, 2) & 2 & 000010 &  & 0 & \\
7 & (4, 1, 3, 6, 5, 2) & 2 & 001101 &  & 0 & \\
8 & (4, 1, 3, 6, 5, 2) & 2 & 000000 & (1521, 1) & 1 & \\
9 & (4, 1, 3, 6, 5, 2) & 2 & 000000 &  & 1 & \\
10 & (4, 1, 3, 6, 5, 2) & 2 & 100000 & (1690, 1) & 2 & \\
11 & (3, 5, 2, 6, 4, 1) & 3 & 001000 & (5005, 3) & 3 & \\
12 & (3, 5, 2, 6, 4, 1) & 3 & 101000 &  & 3 & \\
13 & (3, 5, 2, 6, 4, 1) & 3 & 100001 &  & 3 & \\
14 & (3, 5, 2, 6, 4, 1) & 3 & 001100 &  & 3 & \\
15 & (3, 5, 2, 6, 4, 1) & 3 & 000001 &  & 3 & \\
16 & (1, 4, 2, 6, 5, 3) & 4 & 000010 &  & 3 & \\
17 & (1, 4, 2, 6, 5, 3) & 4 & 000000 & (1625, 1) & 4 & \\
18 & (1, 4, 2, 6, 5, 3) & 4 & 001000 &  & 4 & \\
19 & (1, 4, 2, 6, 5, 3) & 4 & 001000 &  & 4 & \\
20 & (1, 4, 2, 6, 5, 3) & 4 & 100000 &  & 4 & \\
21 & (1, 5, 4, 2, 3, 6) & 5 & 000000 & (1540, 1) & 5 & \\
22 & (1, 5, 4, 2, 3, 6) & 5 & 000000 &  & 5 & \\
23 & (1, 5, 4, 2, 3, 6) & 5 & 100000 &  & 5 & \\
24 & (1, 5, 4, 2, 3, 6) & 5 & 010000 &  & 5 & \\
25 & (1, 5, 4, 2, 3, 6) & 5 & 100000 &  & 5 & \\
26 & (6, 5, 1, 2, 3, 4) & 6 & 000001 &  & 5 & \\
27 & (6, 5, 1, 2, 3, 4) & 6 & 101101 & (41503, 25) & 6 & \\
28 & (6, 5, 1, 2, 3, 4) & 6 & 000011 &  & 6 & \\
29 & (6, 5, 1, 2, 3, 4) & 6 & 100110 & (5775, 4) & 7 & \\
30 & (6, 5, 1, 2, 3, 4) & 6 & 010011 &  & 7 & \\
31 & (5, 4, 2, 3, 1, 6) & 7 & 000100 &  & 7 & \\
32 & (5, 4, 2, 3, 1, 6) & 7 & 001010 & (1375, 1) & 8 & \\
33 & (5, 4, 2, 3, 1, 6) & 7 & 000000 & (1573, 1) & 9 & \\
34 & (5, 4, 2, 3, 1, 6) & 7 & 110000 &  & 9 & \\
35 & (5, 4, 2, 3, 1, 6) & 7 & 100100 & (3185, 2) & 10 & \checkmark\\
36 & (5, 6, 2, 4, 1, 3) & 8 & 010100 &  & 10 & \checkmark\\
37 & (5, 6, 2, 4, 1, 3) & 8 & 100000 &  & 10 & \checkmark\\
38 & (5, 6, 2, 4, 1, 3) & 8 & 100010 & (3125, 2) & 11 & \checkmark\\
39 & (5, 6, 2, 4, 1, 3) & 8 & 011000 &  & 11 & \checkmark\\
40 & (5, 6, 2, 4, 1, 3) & 8 & 011000 &  & 11 & \checkmark\\
41 & (5, 4, 3, 1, 2, 6) & 9 & 011010 &  & 11 & \checkmark\\
42 & (5, 4, 3, 1, 2, 6) & 9 & 001000 &  & 11 & \checkmark\\
43 & (5, 4, 3, 1, 2, 6) & 9 & 000000 & (1617, 1) & 12 & \checkmark\\
\hline
\end{tabular}
\end{center}
\caption{Steps of the factoring.}
\label{table:steps}
\end{table*}

\begin{table*}[t]
\begin{center}
\begin{tabular}{|c|c|c|c|c|c|c|c|c|c|}
\hline
 & Circuit1  & Circuit2  & Circuit3  & Circuit4  & Circuit5  & Circuit6  & Circuit7  & Circuit8  & Circuit9 \\
\hline
$\theta_1$ & -0.619 & 0.190 & -0.513 & -0.619 & -0.867 & -1.667 & 0.400 & -1.133 & 0.476\\
$\theta_2$ & 0.667 & -1.429 & -0.308 & 0.667 & 0.133 & -0.444 & -1.067 & -3.000 & -0.857\\
$\theta_3$ & -1.095 & -0.714 & -1.436 & -1.095 & 0.667 & -0.556 & -0.867 & -1.267 & -1.143\\
$\theta_4$ & -0.095 & -1.381 & -0.205 & -0.095 & 0.067 & 0.333 & -0.933 & -2.067 & -0.095\\
$\theta_5$ & -1.714 & -1.571 & -1.026 & -1.714 & -0.267 & -1.444 & -0.867 & -1.200 & -0.190\\
$\theta_6$ & -0.952 & -2.095 & -0.308 & -0.952 & -0.733 & 0.444 & -0.067 & -1.067 & -0.190\\
$\chi_{12}$ & -0.095 & -0.190 & -0.026 & -0.095 & 0.300 & 0.333 & -0.233 & 0.233 & -0.286\\
$\chi_{13}$ & 0.048 & 0.095 & 0.128 & 0.048 & -0.233 & 0.333 & -0.133 & 0.067 & -0.190\\
$\chi_{14}$ & 0.024 & -0.048 & 0.128 & 0.024 & -0.200 & 0 & -0.033 & 0.033 & -0.238\\
$\chi_{15}$ & 0.048 & -0.024 & 0.103 & 0.048 & -0.067 & 0.278 & 0 & 0.167 & -0.238\\
$\chi_{16}$ & 0.095 & -0.095 & -0.128 & 0.095 & 0.067 & -0.389 & -0.133 & -0.133 & 0.238\\
$\chi_{23}$ & -0.095 & -0.167 & -0.231 & -0.095 & 0.100 & -0.167 & 0.200 & 0.200 & 0.333\\
$\chi_{24}$ & 0.048 & 0.190 & -0.205 & 0.048 & -0.233 & 0.056 & 0.067 & 0.233 & -0.286\\
$\chi_{25}$ & -0.095 & 0.167 & 0.077 & -0.095 & -0.200 & 0.056 & 0.167 & 0.167 & 0.048\\
$\chi_{26}$ & -0.190 & 0.190 & 0.077 & -0.190 & -0.200 & -0.389 & 0 & 0.167 & 0.048\\
$\chi_{34}$ & -0.095 & -0.048 & 0.333 & -0.095 & -0.033 & -0.333 & -0.167 & -0.167 & -0.095\\
$\chi_{35}$ & 0.214 & 0.071 & 0.179 & 0.214 & -0.167 & -0.278 & -0.133 & -0.133 & -0.190\\
$\chi_{36}$ & 0.024 & 0.071 & -0.128 & 0.024 & -0.200 & 0.222 & 0.133 & 0.133 & 0.143\\
$\chi_{45}$ & 0 & -0.119 & -0.128 & 0 & -0.167 & 0 & 0.333 & 0.333 & -0.286\\
$\chi_{46}$ & -0.143 & 0.333 & -0.179 & -0.143 & -0.100 & -0.389 & -0.233 & -0.067 & -0.048\\
$\chi_{56}$ & 0.214 & 0.119 & -0.128 & 0.214 & 0 & -0.444 & -0.100 & -0.267 & -0.286\\
\hline
\end{tabular}
\end{center}
\caption{$R_z$ and ZZ gates rotation angles of quantum circuits used in the factorization of $1591$.}
\label{tab:angles}
\end{table*}

\end{widetext}

\end{document}